# Time-resolved dual-comb spectroscopy with a single electro-optic modulator

Jeong Hyun HUH,[1] Zaijun CHEN,[1] Edoardo VICENTINI,[1,2,3] Theodor W. HÄNSCH,[1,4] and Nathalie PICQUÉ [1,*]

[1]*Max-Planck Institute of Quantum Optics, Hans-Kopfermann-Straße 1, Garching 85748, Germany*
[2]*Dipartimento di Fisica, Politecnico di Milano, Piazza L. Da Vinci, 32 20133 Milano, Italy*
[3]*Istituto di Fotonica e Nanotecnologie, Consiglio Nazionale delle Ricerche, Piazza L. Da Vinci, 32 20133 Milano, Italy*
[4]*Ludwig-Maximilian University of Munich, Faculty of Physics, Schellingstr. 4/III, München 80799, Germany*
*Corresponding author: nathalie.picque@mpq.mpg.de

**Abstract**

**Time-resolved near-infrared absorption spectroscopy of single non-repeatable transient events is performed at high spectral resolution with a dual-comb interferometer using a continuous-wave laser followed by a single electro-optic amplitude modulator. By sharing high-speed electrical/optical components, our spectrometer greatly simplifies the implementation of dual-comb spectroscopy and it offers a high mutual coherence time, measured up to 50-s, without any active stabilization system and/or data processing. The time resolution, which can be reconfigured *a posteriori*, is as short 100 µs in our experimental demonstration. For a span of 36 GHz, the mean signal-to-noise ratio of 80, at 100-MHz spectral resolution and 100-µs measurement time, enables the precise determination of the parameters of rovibrational lines, including intensity or concentration.**

We demonstrate a technique of time-resolved spectrally-multiplexed spectroscopy of single transient events at high spectral resolution and high signal-to-noise ratio, with a time resolution adjustable on the scale of microseconds. Time-resolved spectroscopy of non-repetitive phenomena has a fundamental and technological importance for unravelling photochemical and physical processes that involve transient molecular species. The few spectroscopic techniques that are able to analyze single events face the trade-off between time resolution, spectral resolution, span and signal-to-noise ratio [1, 2]. Such a combination is particularly challenging to implement with multiplex (Fourier transform or dual-comb interferometers) or parallel (dispersive instruments with a camera) spectrometers, which however have the distinguishing advantage that the entire spectrum is measured in a simultaneous fashion rather than in a sequential one. With a multiplex instrument, as all the spectral elements are simultaneously measured on the same photodetector, they are identically affected by the time variations of the sample. Several recent demonstrations of time-resolved frequency-comb spectroscopy have shown an intriguing potential for the investigation of repetitive and reproducible time-varying phenomena [3]. They have been implemented with dispersive spectrometers [4-6] and with dual-comb spectrometers [7-13]. Single-shot time-resolved spectroscopy has also been investigated, in conditions where the signal-to-noise ratio or the spectral resolution were not suited to precise measurements at Doppler-limited resolution [5-9].

We investigate the next step: we develop a dual-comb system based on a single electro-optic modulator that explores the spectroscopy of non-repeatable events with a combination of good signal-to-noise ratio, high spectral resolution (suited to Doppler-broadened profiles of gas-phase small molecules), resolved comb lines and on-demand time-resolution between $10^{-5}$ s and 50 s. A particular focus is to enable the precise determination of line parameters over a large dynamic range of intensities.

Dual-comb spectroscopy exploits the time-domain interference between two optical frequency combs of slightly different repetition rates by automatically and repetitively scanning the optical delay between two asynchronous trains of pulses [3]. In dual-comb spectroscopy and in other techniques of comb-based Fourier transform spectroscopy, the comb line spacing $f_{rep}$ (which provides the spectral resolution for one full interferometric scan), the number of comb lines N and the refresh rate $\Delta f_{rep}$ are fundamentally linked by the relation: $2 N \Delta f_{rep} / f_{rep} \leq 1$. For instance, if N=200 lines spaced by $f_{rep}$=100 MHz are to be resolved, a minimum recording time is $1/\Delta f_{rep} = 4$ μs is necessary. Furthermore, if a high signal-to-noise ratio SNR in the spectrum is required to enable a large dynamic range for the transient phenomenon, acquisition times $T$ longer than the shortest possible $1/\Delta f_{rep}$ may be necessary. From literature results [7, 8, 14-17], a product SNRxNx$T^{-1/2}$ on the order of $10^6$ Hz$^{1/2}$ is a decent figure, especially in conditions where detector nonlinearities are controlled. In such circumstances, a SNR of 100 for a spectrum of N=200 comb lines requires an integration time $T$= 400 μs.

We devise a dual-comb interferometer based a single electro-optic modulator (EOM) which greatly simplifies the implementation of time-resolved absorption dual-comb spectroscopy at high spectral resolution. In the course of this work, we learnt that a similar approach to a dual-comb interferometer based on a single electro-optic modulator had been reported over a narrower span, without in-depth analysis of the applicability to high-resolution spectroscopy [18].

In our set-up, a continuous light-wave at an optical carrier frequency $f_c$ is amplitude modulated by an EOM driven by an electrical temporal waveform $w$ produced by an arbitrary waveform generator (AWG) (Fig.1). The waveform $w(t)$ is a composite signal made of the sum of two multi-tone signals (indexed i=1,2 respectively), each consisting of N cosine waves, starting at an initial offset frequency ($f_{0,i}$) with a frequency spacing increasing of a constant $f_{rep,i}$ from one another:

$$w(t) = \sum_{i=1}^{2} \sum_{n=0}^{N-1} a_{n,i} \cos(2\pi(f_{0,i} + n f_{rep,i})t + \varphi_{n,i})$$

where $a_{n,i}$ and $\varphi_{n,i}$ are the amplitude and phase of the corresponding wave, respectively. The optical spectrum (Fig.2a) is the sum of two combs of slightly different repetition frequencies $f_{rep,1}$ and $f_{rep,2}$. It comprises two spectral bands, centered at $f_c$, with a frequency gap of $2f_0$. In each spectral band, the beat notes between the two combs on a fast photodetector generate a radio-frequency comb of line spacing $\Delta f_{rep} = f_{rep,1} - f_{rep,2}$ starting at frequency $\Delta f_0 = f_{0,1} - f_{0,2}$ set to create enough gap between the two spectral bands to prevent aliasing by optical filtering. The two bands are unambiguously separated by the reflection and transmission output ports of a fiber Bragg grating (FBG).

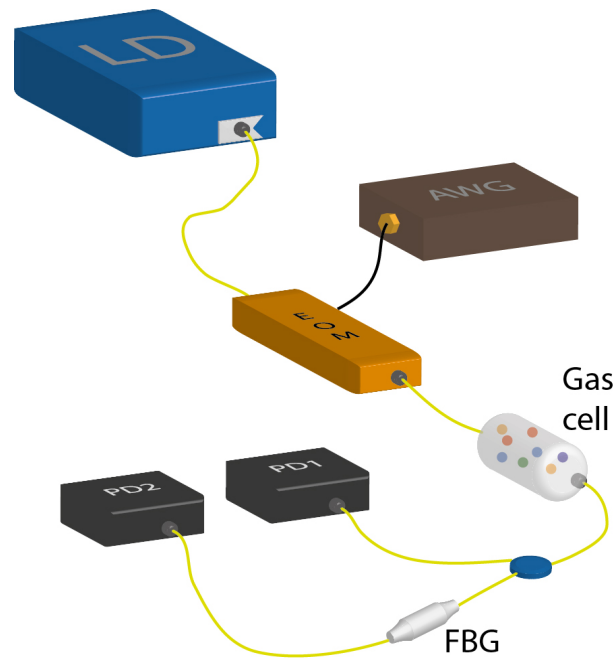

**Figure 1.** Sketch of the experimental principle of single-EO modulator dual-comb spectroscopy. A continuous-wave laser diode (LD) feeds an electro-optic amplitude modulator (EOM) which is driven by an arbitrary waveform generator (AWG). The modulated beam interrogates the gas sample. The two spectral bands that are aliased in the radio-frequency domain are separated using a fiber Bragg grating (FBG) and their time-domain interference is measured on fast photodetectors (PD1 and PD2).

The design strategies of the multi-tone modulation signal influence the overall performance of the interferometer. In particular the phases of the cosine waves relative to one another turn out pivotal for achieving a good signal-to-noise ratio.

A constant or linear temporal phase distribution is not favorable (Fig. 2b), for the phase of all frequency components are aligned at given times multiple of $2\pi$ and their constructive interference forms radio-frequency pulses. Owing to the high peak-to-average power ratio (PAPR) of a radio-frequency pulse, the typical noise characteristics of high-bandwidth AWGs lead to a relatively low SNR per radio-frequency comb line. Moreover, for operating the EOM within its linear regime, the peak amplitude of the input modulation signals (i.e. the radio-frequency pulses) should be low enough to satisfy the small-angle approximation for the sinusoidal transmission curve of the EO modulator ($\sin(x) \sim x$). This limits the dynamic range.

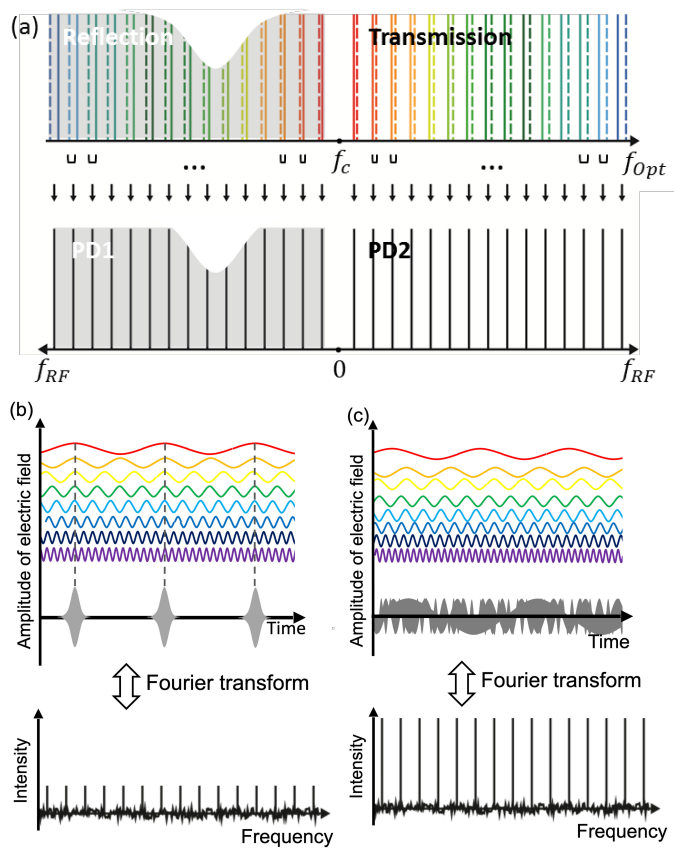

**Figure 2. (a)** Frequency domain picture of dual-comb spectroscopy with a single EO modulator: two optical frequency combs of slightly different line spacing, centered at the frequency $f_c$, interrogate an absorption line. Each side band is separated by a FBG and their beating signal generates a radio-frequency spectrum; **(b),(c)** Time domain and frequency domain comparison of multi-tone signals with (b) linear phase distribution and (c) quadratic phase distribution (c).

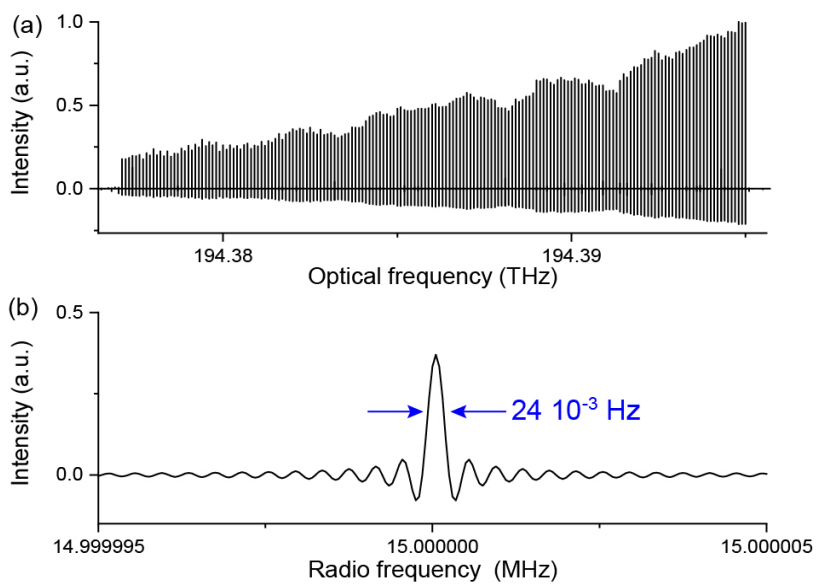

**Figure 3. (a)** Experimental dual-comb spectrum resulting from the Fourier transform of a 50-s measurement. The spectrum contains 180 comb lines over a 18-GHz span at 100-MHz line spacing **(b)** Magnification of a portion of (a) showing a single comb line with its cardinal-sine instrumental line shape.

Unlike a mode-locked laser, our electro-optic comb generator allows complete freedom in tailoring the temporal phases of the individual comb lines. A multi-tone waveform with a better peak-to-average power ratio may be obtained by using a quadratic phase profile, i.e., $\varphi_n \propto n^2$ (Fig. 2c). For example, with N=180 waves and a 100-MHz line spacing, the measured peak-to-average power ratio of a quadratic profile is 5 dB lower than that of a random profile, and 20 dB lower than that a linear profile, resulting in a SNR per radio-frequency comb line higher by the same factors. An additional optimization algorithm could be employed to calculate a phase profile with a lower PAPR based on the quadratic phase profile as an educated guess [19].

In our experimental implementation, a tunable continuous-wave laser diode emits at a frequency $f_c$ between 183 and 198 THz. Its beam is amplitude modulated by an EOM driven by the combined multi-tone signals. The amplitude modulator is biased to operate near the minimum of transmitted power, so that the transmitted light field depends linearly on the driving signal. Each multi-tone radio-frequency signal consists of N=180 waves with a quadratic phase profile. The two repetition frequencies $f_{rep,1}$ and $f_{rep,2}$ can be freely adjusted over a large range, from 1 MHz to 10 GHz. The smallest possible difference in line spacing $\Delta f_{rep}$ is 32 Hz, set by the ratio of the sampling rate (64 $10^9$ sample s$^{-1}$) to the maximum waveform memory (2 $10^9$ samples) of the AWG.

In the illustrations given in this Letter, the frequency of the continuous-wave laser is 194.399 THz. The line spacing are chosen to $f_{rep,1}$ = 100 MHz and $f_{rep,2}$ = 100.01 MHz, respectively and the offset frequencies are $f_{0,1}$= 4.005 GHz and $f_{0,2}$= 4.006 GHz, respectively. The driving signal produced by the AWG comprises 6.4 $10^6$ samples at a rate of 64 $10^9$ sample s$^{-1}$ and a vertical resolution of 8 bits. The optical signal spans two bands of 18 GHz separated by a 8-GHz gap. Time-domain interferograms recur at a rate of 10 kHz and are measured by two avalanche photo-detectors. The electric interference signal is sampled at $10^7$ sample s$^{-1}$ by a 18-bit digitizer and it is Fourier transformed. Figure 3(a) shows an experimental dual-comb spectrum, Fourier transform of an interferometric signal recorded within 50 seconds. The duration of the recording is limited by our data acquisition capabilities. The spectrum is calculated using a complex Fourier transform, without any data processing, and its amplitude is shown in Fig. 3(a) on its optical-frequency scale. The spectrum with 180 resolved comb lines across 18 GHz corresponds to the frequency band reflected by the FBG. The full-width at half-maximum of the individual comb lines is 24 $10^{-3}$ Hz in the radio-frequency scale (Fig.3(b)). The shape of each line appears as a cardinal sine, signature of the finite measurement time of the time-domain interferogram. Observing such a line shape, identical to the theoretical expectation, shows that our interferometer keeps coherence for more than 50 seconds.

Electro-optic modulators have been broadly used for dual-comb spectroscopy [10, 14-18, 20-25]. Although the spectral span of such frequency-agile combs and the achievable frequency accuracy in the spectra are highly limited compared to those of fiber lasers, EOMs represent an easy-to-use alternative, where line spacing and spectral position can be quickly and freely selected by simply dialing a knob. Our development brings further simplification by using a single modulator rather than two. The coherence time of the interferometer is much enhanced, here to 50 seconds (limited by our data acquisition capabilities), compared to that of an interferometer comprising two arms of electro-optics, which culminates at 1 second [24] in the same laboratory environment. In both cases, such a coherence time is obtained entirely passively, without any servo-control system nor analog or digital processing. The interferometer based on a single EOM dramatically simplifies the implementation of dual-comb spectroscopy. Owing to the long coherence time, the signal, which can be sampled continuously, leads to a highly consistent data set. In this example, the interferograms recur every 1000 time-samples exactly, during 50 seconds, facilitating reconfiguration of the temporal resolution and/or real-time averaging without any corrections of data.

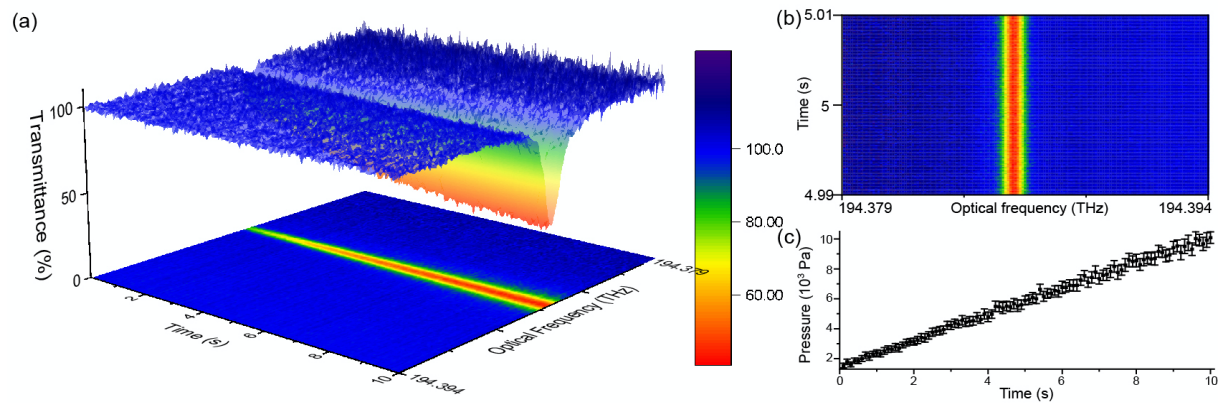

**Figure 4. (a)** Evolution of the dual-comb absorption spectrum of the $P(27)$ line of the $n_1+n_3$ band of $^{12}C_2H_2$ over a 10-s time window. The spectral span is 15 GHz, the spectral resolution is 100 MHz and the time resolution is 100 μs. **(b)** Expanded portion of (a). showing an intensity map of a 20-ms time window at around 5 s, containing 2000 spectra. **(c)** Time-resolved pressure of $^{12}C_2H_2$ retrieved from a fit of the spectra in (a). The error bars are the 95% confidence intervals (2 standard deviations).

Next, we explore the potential of the spectrometer for time-resolved spectroscopy of a non-repetitive event. A fiber-coupled cell, of a length of 16.7 cm, is initially under primary vacuum. Within 50 seconds, the cell is filled with acetylene up to a pressure of about 2.5 $10^4$ Pa, while the time-domain dual-comb interference is simultaneously sampled in a continuous way.

Figure 4(a) shows the time-resolved absorption spectra of the $P(27)$ line of the $\nu_1+\nu_3$ band in $^{12}C_2H_2$, at 194.38633 THz, evolving over a 10-s time window at 100-μs time resolution, as the gas pressure increases. The 10-s time window is selected out of the total 50-s acquisition time where the line profile changes most rapidly. A reference spectrum for baseline calculation is obtained from a spectrum averaged over 1 s without the cell. The frequency resolution is 100 MHz over the 15-GHz span corresponding to the spectrum of the lower frequency band, reflected from the FBG. The mean SNR of the transmittance baseline is as high as 81 for a 100-μs measurement with an optical power of 3.3 μW. The SNR is defined as the transmittance signal divided by the standard deviation of the noise on the baseline. The low optical power at the detector is due to the limited peak-to-peak voltage of the driving signal from the AWG. The power could be increased by using a linear wide-band amplifier at the output of the AWG or by using an optical amplifier at the output of the EO modulator. Nevertheless, the figure-of-merit corresponding to our measurement conditions, defined as the SNRxNxT$^{-1/2}$, is 1.5 $10^6$ Hz$^{1/2}$, and it compares favorably with that reported in other implementations involving electro-optic combs in the telecommunication region [14-16]. In our experiment, the figure of merit is detector-noise limited. Figure 4(b) shows a magnified image for a 20-ms time window around 5 s comprising 2000 spectra. Within the short measurement time, the transmittance seems constant as the pressure variation is negligible. Figure 4(c) presents, over a 10-s time window, estimated values of pressure (only 100 values out of 100,000 data) varying from 1.55 kPa to 10.16 kPa and their 95% confidence intervals (plotted as error bars), which correspond to two standard deviations. The pressures are extracted from the nonlinear least-square fit of a Voigt profile, using the line intensity available in the HITRAN database. Relative errors are within 10 % and the relative errors at low pressures are generally larger due to the lower absorption depths.

The time resolution can be chosen *a posteriori*. In our demonstration, it can take on any value between $1/\Delta f_{rep}$ = 100 µs and 50 s. This will prove useful for time-resolved spectroscopy of single events, where the variation of the observed physico-chemical phenomena may be *a priori* unknown. With our technique, the time resolution can be reconfigured arbitrarily and the trade-off between integration time and signal-to-noise ratio can be straightforwardly adjusted to the scientific objectives. While shorter time-resolution for the spectroscopy of non-repetitive phenomena can be reached [1, 2], our technique shows an unmatched compromise between spectral resolution, signal-to-noise ratio, span and time resolution, which enables precise determination of all line parameters, necessary for quantitative measurements.

While the spectral span is limited by the capabilities of opto-electronics systems, it is possible to spectrally tailor the interrogation by feeding the electro-optic modulator with several continuous-wave lasers [17]. Furthermore, other spectral regions, such as the mid-infrared fingerprint region, may be reached by nonlinear frequency conversion [10, 24]. Interestingly, our technique shows an intriguing potential for a high-resolution spectrometer on a photonic chip using thin-film lithium niobate integrated modulators [26], which might enable scalable fabrication techniques and the integration of the spectrometer sub-component into more complex photonics functionalities towards a spectroscopy laboratory on a chip. Our technique thus holds promise for a miniaturized field instrument and opens up new possibilities for real-time spectroscopic diagnostics and sensing.

**Funding.** Max-Planck Society. Max-Planck Fraunhofer cooperations. Carl-Friedrich von Siemens Foundation.